# IT Project Showstopper Framework: the View of Practitioners

Godfred Yaw Koi-Akrofi

Department of IT Studies, University of Professional Studies, Accra

## Abstract

*The study intended to unravel critical IT project showstoppers which tend to halt IT projects temporarily or permanently, and ultimately cause them to fail, by positioning them in the systems development life cycle (SDLC) framework. Interviewing 8 IT project and program managers of the banking and telecommunications industries in Ghana individually and in a group, 19 critical showstoppers were identified spanning the whole SDLC. Generally, it was observed that for the successful completion of IT projects, the expertise and availability of project managers and team members are critical. Again, the project manager must be able to prove that the project is in line with the objectives and strategic direction of the business, is being mounted to gain competitive advantage, and has a solid business case. Thirdly, funding is key at all stages of the cycle, as well as approval for continuation at various stages.*

## 1. Introduction

IT has taken over businesses today due to its enormous benefits to the same in terms of profitability and performance in general. This has necessitated businesses worldwide to invest massively in IT infrastructure and software. Investments in IT are accomplished through projects, and IT projects, unlike other projects, are very difficult to complete and troublesome due to the behavior of the systems involved and the behavior of the intended users. This makes the failure rate for IT projects higher than all other projects. According to the CHAOS study which was published in 1995 by the Standish Group, 31% of all IT projects were canceled before completion, and 53 % of IT projects were completed over budget/schedule which was also referred to as challenged projects; they did not meet all of the project requirements [1]. The author has done extensive work on the failure of IT projects worth mentioning to support this work. In his work "Delivering successful IT projects: A literature-based framework", he came out with four key critical success factors without which IT projects are bound to fail. These are excellent skills of project managers/team members, positive top management involvement, proper methodology/processes and governance structures in place, and good communication [2]. In other research, the author in collaboration with others, established the need for key project control variables to be managed and balanced well to ensure project success [3]. Again, in their paper "IT project failures in organizations in Ghana", they came out with seven common causes of failure of IT projects namely: lack of project management departments, no quality checks for IT projects, IT projects not completed according to schedule, no specific IT project management methodology followed, revision of scope very often for a particular project, IT projects not meeting users' needs, and wrong estimates [4]. IT project governance was also reported by Aon to be the majority cause of a list of causes of project failure; the office of Government commerce of the UK Government in conjunction with the National Audit came out with a guideline in 2007 which lists eight causes of project failures. Six out of the eight causes were attributed to governance issues [5]. In a related study, other researchers grouped IT project failures into four categories: Correspondence failure (the information system (IS) fails to meet its design and objectives), Process failure (the IS overruns its budget and/or time constraints), Interaction failure (the users maintain low or no interaction with the IS), and Expectation failure (the IS does not meet stakeholders' expectations) ([6], [7], [8]). All the causes of failure for IT projects mentioned in the studies above include failures resulting from showstoppers and projects that are considered failures even though the projects were completed.

IT project failure engulfs project showstoppers. A showstopper can lead to the outright cancellation of the project. It can also affect the time allocated for the project if not resolved quickly. In summary, a showstopper will not be a problem if it is resolved quickly and does not affect the time allocated for the project, otherwise, the project will be deemed a failure, even though it goes onto completion, for the fact of going beyond the allocated time for the completion of the project. IT project failure is broader in the sense that even if a project goes to completion and is completed beyond the budget and time, it is still considered a failure or unsuccessful. The same applies to a project completed with the output not meeting the users' specifications [9]. Project showstopper comes in when the project is halted along the way, and there is no deliverable to show forth as the result of the project. In a nutshell, a project showstopper has to do with anything that has the potential to halt a project permanently (cancellation) or temporarily (issue/s resolved, and the project goes onto completion later). At every stage of the project management life cycle, there are critical issues that must be tackled to ensure the continuation of the project. For a typical systems development life cycle, what is popularly known as Waterfall, the next phase can never start without the completion of the current phase with documentation. The case for agility is different. This study is situated in a typical traditional systems development methodology like the waterfall. Agility has taken over from SDLC in industry, especially for Systems/software development, but for all other IT projects apart from Systems/software development, SDLC is still being used. This study is situated in IT projects outside the scope of Systems/software development, where SDLC is still being employed. Figure 1 below shows how the author conceptualizes what constitutes IT project failure.

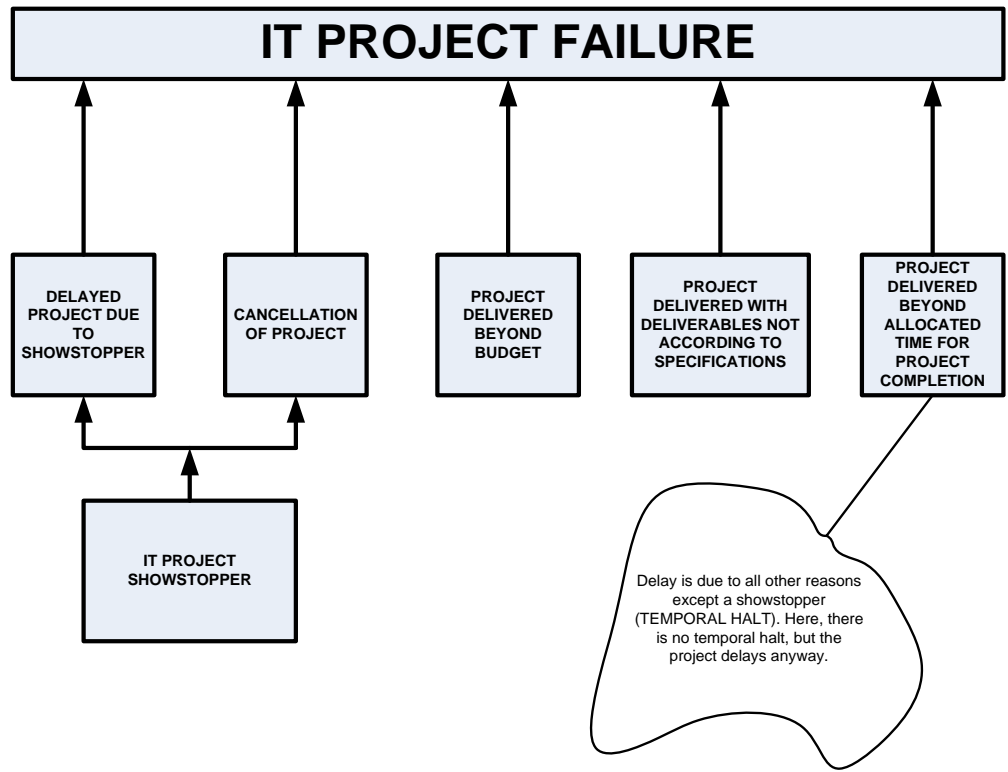

**Figure 1:** Conceptual Framework

**Source:** Author

### 1.1 The objectives of the study

The objective of this study is to contribute to the general body of knowledge and research work in the causes of IT projects. Specifically, the work is aimed at exploring the critical showstoppers of IT projects for a typical traditional systems development methodology.

### 1.2 The research Problem

IT project practitioners are always quick to state the three reasons for going beyond stipulated project completion time, over budget and deliverables not according to specifications as reasons for project failure. They normally lose sight of the critical showstoppers which in most cases halt the projects temporarily or permanently, and ultimately cause IT projects to fail. This study intends to unravel some of these showstoppers and place them in a framework of the systems development life cycle (SDLC), to help practitioners curb occurrences of IT project failure due to showstoppers.

## 2. LITERATURE REVIEW

Information Technology Project failure has many dimensions. According to Dmitriy and Mayur [10], adopting a high proportion of risky projects can lead to a high number of project failures. A risk is an anticipated danger in the delivery of a project. This can halt a project. A good project manager is one who can predict risks and mitigate them, so they do not hamper the progress of the project. In a bucket of related projects as in program management, the program manager should be able to know which projects are of high risks and which are not.

Whittaker [11] found out that there are three main reasons for IT project failure namely poor project planning, a weak business case, and a lack of top management involvement and support. Without good planning, there is nothing good to deliver. Good implementation is based on good planning, and so poor planning will affect the output or deliverable negatively in terms of quality; deliverables will not be according to specifications. Ibrahim, Ayazi, Nasrmalek, and Nakhat [12] also mentioned top management support and involvement as one of the critical things that can ensure the success of IT projects. Doing a literature review on the subject matter of IT project failures, Ibrahim, Ayazi, Nasrmalek, and Nakhat [12] maintained that apart from the top management support and involvement, other three critical factors can ensure the success of IT projects, and these are, good consultant effectiveness, good project management effectiveness, and proper User Involvement.

According to Bannerman [9], a project is successful when it is completed within time, within budget, and when deliverables are according to specifications. This definition is generic. Other researchers also have other views on the same subject matter. Guiney [13] defined project success from two levels. The level one has to do with all the components in that of Bannerman [9] plus the project delivered on the agreed scope. The scope dimension is the only addition. The second level has to do with the dependencies which he named as people management factors, project design and definition factors, and organizational support factors. These dependencies are the same for the four reasons for project success at level one. Guiney [13] further explained that People management factors include all factors related to creating and maintaining effective team members and Project design and definition factors include all factors related to creating and maintaining a project environment supportive to the team, whereas Organisational support factors include all factors related to creating and maintaining organizational processes that support the project team.

On the other hand, looking at the failure definition for IT projects exposes other components that broaden the scope of the IT project failure-success discussion. One of such is the concept of IT project showstoppers. From the concept in figure 1, we realize that project cancelation is a major cause of project failure. Once the project is cancelled, it has automatically failed, and this is a major project showstopper. This work details some of the reasons within the project life cycle that can cause cancelation, and hence present the project as a failure.

Again, some of the showstoppers will only halt the project for a while and continued later when the showstoppers are removed. This may result in the overall delay of the project. Other factors apart from the commonly known showstoppers can also delay IT projects. For example, control variables interdependencies may result in the delay of a project. Sheer laziness on the part of team members may even delay a project. Cost and scope are variables that must be balanced to ensure project success, as they may pose as showstoppers during the project delivery. Compliance with specifications must be monitored closely from the beginning to the end to ensure that deliverables are according to specifications. This falls under quality management. Non-compliance to specifications if not detected earlier, will only be detected when the deliverable is deployed for use and the clients begin to complain.

In summary, the variables critical for IT project success are time, scope, budget (cost), and quality. Once these variables are not managed well, the IT project may fail.

Guiney [13] is of the view that the underlying factors to IT project success (the second level to the definition) have to do with people management factors, project design, and definition factors and organizational support factors. To Guiney [13], once these factors are gotten right, they will influence the first level factors of time, scope, budget (cost), and deliverables according to specifications (quality) to produce IT project success.

This work looks at IT project failures, not from the variables discussed earlier, but from anything that can halt a project temporarily or permanently and can go-ahead to render the project a failure. If the showstopper results in the cancelation of the project, the project automatically has failed. If the showstopper results in the halt of the project and is continued later, the project may not be a failure if it is completed on time, and all the other conditions are met. Otherwise, it will still be considered a failure. This work is a creation of the views of experts in this space. The Systems Development Life Cycle is the common traditional methodology employed for IT projects, and this work employs that to fish out the common IT project showstoppers that have the propensity to cause IT project failures. These showstoppers may take any shape and form; may be linked to any control variable described earlier. The most important thing is that they are factors identified by the experts as potentials for project failures.

## 3. MATERIALS AND METHODS

### 3.1 Research Method

This work solely adopts a qualitative approach. IT system, program, and project managers were interviewed to offer their candid views on the subject. The information collected from these interviews and discussions formed the basis for the study and creation of the framework. A quantitative approach is not an option here because there are no parameters to determine relationships, and so on. The outcome of this qualitative study will bring out some of the parameters in this space to set the tone for quantitative analysis.

### 3.2 Research Type

The research format is mainly Exploratory. Exploratory, because primary research was used as an exploratory method. Experienced IT program and project managers were interviewed on the subject individually and as a group for an experienced survey.

### 3.3 Techniques/tools/approaches/instrumentation/devices

Structured questions were asked separately and, in a group, to the IT project and program managers. For the individual interview as well as the group interview, a discussion format was used. Notes were taken during the discussions and recordings were made on mobile phones.

### 3.4 Data Collection Methods

The study case was the service sector, primarily the Telecommunications and the banking industries in Ghana, and primary data was solely used for the analysis. No secondary data was used. The service sector in Ghana was used because they have within their setups project management departments and the expertise for this kind of study. Again, the service sector is known to be one of the sectors in Ghana that invests so much in IT.

### 3.5 Population and sampling procedures

The study employed non-probability sampling methods of quota and convenience. Quota in the sense that representative individuals were chosen from a subgroup as against random sampling of members of the group. In this instance, the study focused on the IT program/project managers as representatives from the project management offices (PMOs), which are units/subgroups under the Technology departments of the companies in the telecom and banking industries in Ghana. The reason for the researcher was that the subject of the research is a specialized area that may be best understood by practitioners, and not anybody working in the PMO unit or Technology department. It is also convenient because the industries in the sector were easily accessible to the researcher and easier to reach in terms of proximity. In summary, the population employed was IT project and program managers of the Telcos and banking industries in Ghana, and the sample size was 2 IT project managers and 2 program managers for each industry: Telecommunication and banking, making up to 8 people.

## 4. RESULTS

The basic question asked the practitioners individually and in a group at different points in time was 'what are the major showstoppers for an IT project for a typical SDLC?' The typical project management process groups are Initiation, Planning, Executing, Monitoring and Controlling, and Closing. This is generic, and so when put side by side with the SDLC or the Waterfall, we have Initiation as Analysis, Planning as Design, Executing, Monitoring and Controlling as Implementation or Development or Coding, and Closing as Testing, Go-live, and Deployment.
The responses from the experts to the basic question asked are summarized in the ensuing write-up and put into a framework later at the end of this write-up.

Practically, in industry practice, the experts maintained that the Initiation process group is made up of two key processes: the project concept (Idea formulation) stage and the Feasibility Stage. Anyone in the business can drop an idea for consideration and conversion into a project. It does not matter where it is coming from. It can be from the top (from management) or the bottom (from operations). At this stage, it is just an idea, and all things being equal, this idea will only survive if it is in line with the business strategy or strategic plan or business objectives. Any idea of a project not in line with the business objectives of an organization will immediately be shot down, regardless of the source of the idea. Unanimously, it was therefore, agreed that at the concept stage, the main showstopper is an idea of a project that is not in line with the objectives of the business.

Still under initiation, if the idea of a project passes the test of the business objectives, the idea enters the feasibility stage. Here we are talking about the feasibility item to determine and the determination itself. For example, the feasibility item to determine may be technical feasibility (whether technically, the project can be done, and whether there is technology in place to support it), but it does not end there. To be able to get the outcome of feasibility, that is, whether the project can proceed or not, you need to commit resources to determine it. At this stage, many showstoppers can either stop the project temporarily or permanently. Some of the showstoppers that were agreed upon are discussed below:

- *Insufficient customer requirements:* Inadequate information on user requirements is a major showstopper at the feasibility stage. Lack of information on user requirements is a non-starter. The IT system is being built or developed for users, and so if their requirements are not taken or captured and factored into the design and the development, how are the users going to appreciate the final output? The system may become a 'white elephant' in the end, with no one patronizing it. Users may go ahead to reject the output outright. To avoid this scenario, it is better to get full user requirements before proceeding further. Insufficient requirements may be caused by client-developer communication failures. The technical language of the developer may not be understood by the client, and this usually prevents the clients from coming out clearly, and this results in the developer not getting the adequate requirement needed. Another problem has to do with resistance to change on the part of the clients. If the clients have a fair knowledge of the project, and they see it as a border with the explanation that they are okay with the current system or software, the probability of they opposing the project will be very high, and this can cause them not to be forthcoming with requirements. Indifference or apathy can also cause clients not to be forthcoming with requirements. Organizational politics can play a role here too, especially when the clients or customers are internal. Top managers who are not in favor of the project can influence the intended users whom requirements will be taken from, frustrate the efforts of the developer in terms of requirement gathering, to kill the project right at this stage.

- *Lack of funding for feasibility:* You need funding to perform feasibility. Once the business is reluctant to provide funding, the project cannot proceed, because the risks involved in proceeding without performing feasibility may be damning, and no reasonable project manager would want to do that. This makes a lack of funding for feasibility a major showstopper at this stage. Feasibility is critical, especially in uncharted waters, and in situations where the fundamental parameters have changed. Feasibility has many dimensions: Technical, operational, financial, economic, and so on. Technically, the business may not have in-house resources to check feasibility. Taking a radio access network (RAN) rollout project as an example, the business would have to go out and engage vendors or contractors to do the feasibility if there are no experts in-house to do it. This requires money because these contractors would have to be paid. This cost is not part of the cost of the rollout itself. Financial feasibility is key because the business will need an expert to investigate the financials based on the prevailing macroeconomic parameters and predict the cost of the project. This financial expert would have to be paid if the business has not gotten one.

- *Lack of expertise (Human Resources) for feasibility:* lack of the required expertise for feasibility can also be a major showstopper. For example, a project requiring the Customer Relationship Marketing (CRM) tool may require an expert to do feasibility for the requirement for a customized version apart from the base version procured. Based on this feasibility, the customized version can be developed fully for operation. Lack of an expert for this can be a major showstopper for the CRM project. In this case you may have the money at hand to pay, but the business's inability to get an expert may be a showstopper.

- *The timing of feasibility:* The business priorities can cripple the feasibility. The business can deprioritize the feasibility depending on the ongoing business roadmap and cause the project to halt. This may be due to the engagement of resources for other important things in the business. The very people who will do the feasibility may be engaged in other operational or project duties. The line in the budget earmarked for the feasibility may be collapsed and used for other equally important and pressing activities. Once feasibility is not done, the project cannot be continued.

- *Denial of approval based on the business case:* A bad business case can throw the project off. A viable business case with good returns (tangible and intangible) will stand. If the business case developed for the project is not convincing based on agreed indicators, the project will be discontinued. A good business case will pass the project to be continued. Approval will then be given, but can still be discontinued, even after approval, if there is no adequate funding. A typical example is when the benefits are not clear or when the cost of the project far outweighs the benefits even for years after the delivered project has been in operation. The trend now is to look for more intangible benefits as against the tangible benefits. The intangibles are not supported with financial ratios and figures but may later produce tangible benefits. In some jurisdictions, this is also known as cost-benefit analysis where project evaluation methods like payback period, internal rate of return, net present value, return on investment, and so on are employed for tangible benefits determination for capital budgeting decisions.

Having gone through the initiation process and secured approval and funding for the project, the project enters the planning stage. This is where detailed plans are made for all spheres of the implementation process. At the planning stage, the main showstopper that can be encountered is a lack of experts to do the planning. Smooth execution is based on proper planning. Without expert hands to plan, there is nothing to execute, and the project can be halted. The key resource in planning or designing is the human resource; once we get experts who can plan or design, the planning stage will be smooth.

After planning or designing, the stage is set for the actual business, implementation, or execution. This is the stage where the actual work is done physically. Planning is mostly conceptual and logical, but the physical set up takes place at the implementation stage. A lot of the problems in project management are encountered at this stage because it is at this stage that all energies and strategies are marshaled to deliver the main deliverable. Many things can serve as showstoppers at this stage, and they are discussed as follows:

- *A shift in business focus:* It does not matter all that has been done to this level, if the business focus shifts, the project can be halted and eventually be canceled. Business focus shifting means that the project now is not in line with the revised or current objectives of the business defining the new focus. Business objectives may change due to the actions of competitors in the industry, the macroeconomic landscape of the country, the behavioral pattern of the customers, the financials of the business, and so on. These and other factors are indicators that can change the focus of the business, all in the bid to ensure survival and business continuity. If management thinks that the success of a new focus is not dependent on the project at stake, the project may be halted immediately.

- *Losing key resources on the project:* One of the key attributes of a project is that it consumes resources. Projects are delivered using key resources, including human resources. The technical team in the case of IT projects delivers the project. Losing key resources is a disaster to any project team and can lead to the halt or outright cancellation of the project. This is because the people may be irreplaceable in terms of knowledge and skills in the interim. Key human resources can be lost due to competitors poaching them with the main aim to delay or halt the project outright, to gain an advantage, especially if the project has a lot of prospects. Poor remuneration can also cause key resources to leave. Experts of this sort are kept at bay with nice retention packages.

- *A massive change in business requirements:* Massive change in business requirements will change scope drastically, and the project may not survive going forward, as the change may affect other project control variables or indicators such as cost, time, quality, benefits, and risks. In the agile environment, massive change in business requirements may not be an issue, but in an SDLC environment, it is a big issue. This is because, for SDLC, the next phase cannot be started without completing the current phase; it is a strict and ordered methodology. Once the requirements gathering and analysis phase is completed, it is difficult coming back to redo. It will mean starting all over again, which is going to affect timelines, costs, and other resources. These are the very things that can trigger a halt.

- *Supplier challenges in delivering software or hardware:* This can also be a major showstopper as the software and hardware may be needed during the project delivery. The project may be such that progress cannot be made at a point in time without this software and hardware at hand. Sometimes the supplier may be required to do customization for the organization in question. This may require a complete project at the supplier side as the customization may include requirement gathering from the organization, analysis, design, and construction before the final product is delivered to the organization. Sometimes too, if for instance, it has to do with hardware delivery, the time for shipment and clearing from the port may also be an issue. The delay may be more pronounced when there is a pandemic like Covid-19 lockdown where there are restrictions on movements at the ports and so on.

- *Regulatory demands:* Regulatory bodies like the National Communications Authority (NCA) in the case of the Telcos, The Bank of Ghana (BoG) in the case of the banks and for some aspects of Telcos (for example, mobile money projects) and the Ministry of Communications can make demands in the course of the project, and cause the project to cease temporarily or permanently. Before the start of the project, the policy of the regulatory body may have not even been conceived, only to be conceived and enforced later when the project is far advanced, and then the project forced to stop as a result of the policy.

- *Force Majeure:* A natural disaster can occur and put a stop to a hitherto ongoing project. Risks of this nature may be avoided in the IT space if hardware and software are properly secured. In other instances, if nothing can be done, then the project would have to stop or halt for some time.

- *Cyber Attacks on Systems:* Cyber-attacks can also cause projects to halt temporarily or permanently. They are a major showstopper in IT projects. Systems can be ceased by external aggressors, and this can prevent the team from working until they get access to the systems. Once systems are compromised, it is better to stop the project and find better ways of securing them before continuing the project. Sometimes the system would have to be changed completely to outwit the hackers. This will mean that the project has to be halted for some time.

- *Business case no more viable due to competition action:* Competitors in the same industry can outsmart each other to the detriment of the competitor's plans to gain so much from their novel and intended cash cow projects in the industry market share. This behavior defeats the business case for the projects of the affected company, and forces management to discontinue the projects. It is an executive decision and so cannot be taken by lower managers. Again, the benefits of the project may not be applicable at a point in time during the project delivery. This may cause management to halt the project, especially knowing very well that benefits no longer hold, and the fact that the business would have to spend so much in terms of resources to complete the project.

After the successful completion of the IT project, it enters the testing phase. The testing phase prepares the developed system for go-live and eventual deployment to intended users. The testing is to ensure that there are no issues (errors and defects) unresolved and that the system is fully healthy and ready for use by the intended users. During testing, two main showstoppers can occur according to the experts, and they are discussed as follows:

- *Failed System Integration Test (SIT), User Acceptance Test (UAT) test:* In most cases, industry-agreed and acceptance percentage for SIT is 99% and beyond, and that for UAT is 90% and beyond. Anything less than this is a failure, which calls for investigations and resolution of issues. Once SIT and UAT fails and issues are unresolved, there is no way the project team can go-live. This can be a major showstopper. All issues must be resolved and all items on the SIT and UAT lists be ticked as cleared before go-live can go on or happen. To avoid this delay in the case of software errors and their resolution during SIT and UAT, developers and the testing team more often rush through the process to give clearance and sign off the SIT and UAT documents for deployment only to find later that due diligence was not done when errors graduate into faults and system failures.

- *Lack of skilled UAT, SIT adequate testers, and materials:* Two issues are presented here: skills of testers and appropriate materials for testing (testing kit). The two must be in place for effective testing. If one is lacking, the testing cannot go on as planned, and this can cause a project to halt. If the testers do not have the requisite skills for the testing exercise, they may give clearance without touching on the salient things, therefore, hiding some errors. In the same vein, if the right testing kit is not used, some errors may not be detected.

Once testing is successful, the project team is ready to go-live and then finally deploy to intended users. The go-live experience is a mixed one: a combination of excitement that the project is coming to an end, and that the team's efforts are being rewarded, and a feeling of jittery not too sure if all is well up to that point. At this stage, nothing should be given to chance. All loopholes must be closed to ensure a successful go-live. Regardless of all these, something can go wrong to halt the project from going live. It should be clear that an IT project is not completed until the team goes live successfully. According to the experts, two main things can occur to halt the go-live experience. These are discussed as follows:

- *Live deployment change request decline:* The team may be fully ready for go-live, but the approval for go-live based on Request For Change (RFC) may be approved or disapproved by the change control team based on several factors. The change control team may consider several factors such as customers on the life system who may be affected or impacted negatively due to the period or timing of the go-live, the ambiguity in rollback plan in case there is a problem, the revenue that will be lost, and so on. In most cases, the change control team may not risk these factors to go ahead and give approval, and this can be a showstopper to the project. Practically, it can get to a point where all approval signatories to the change request form may decline if there are high anticipated risks and nobody wants to take responsibility in the event something untoward happens.

- *Improver or no proper operational readiness:* It is one thing for the team to be ready for live deployment and another thing for operations to be ready to host and maintain the IT system. Hosting and maintaining a system go with a lot of responsibility including human resources, training, and mental readiness. Sometimes there must be in place a change management team to help intended users appreciate the impending change in software and hardware systems. If these things are overlooked, the go-live can occur anyway, but the people may not patronize the system afterward. If the operational staff is not trained on the system, how are they going to use it? These are all preparations that must be made before live deployment. If these preparations are not made, it will be better to hold on to live deployment to forestall post-deployment problems. In summary, the non-readiness of operational staff can halt live deployment.

International Journal of Software Engineering & Applications (IJSEA), Vol.11, No.3, May 2020

Once the live deployment is successful, the project officially ends, and the system is handed over to operations for hosting and maintenance. The project team is dissolved, resources are released, and the cycle begins again for another project. A pictorial IT project showstopper framework (ITPSF) based on the discussion above is shown in figure 2 below. The first column of figure 2 is the generic project management process, fashioned according to the project management institute's (PMI's) project management process group. This is matched to the SDLC or the Waterfall which is calved out of the generic one purposely for IT traditional projects. This is again matched to the process that is prevalent in the industry, at least the two industries under consideration (the Telcos and the banking industries) as was revealed by the experts engaged. The last column contains the revealed IT project showstoppers for each phase of the process. The figure presents a holistic view for a glance of the various showstoppers for each phase of the traditional project management methodology or process irrespective of the direction one views it from.

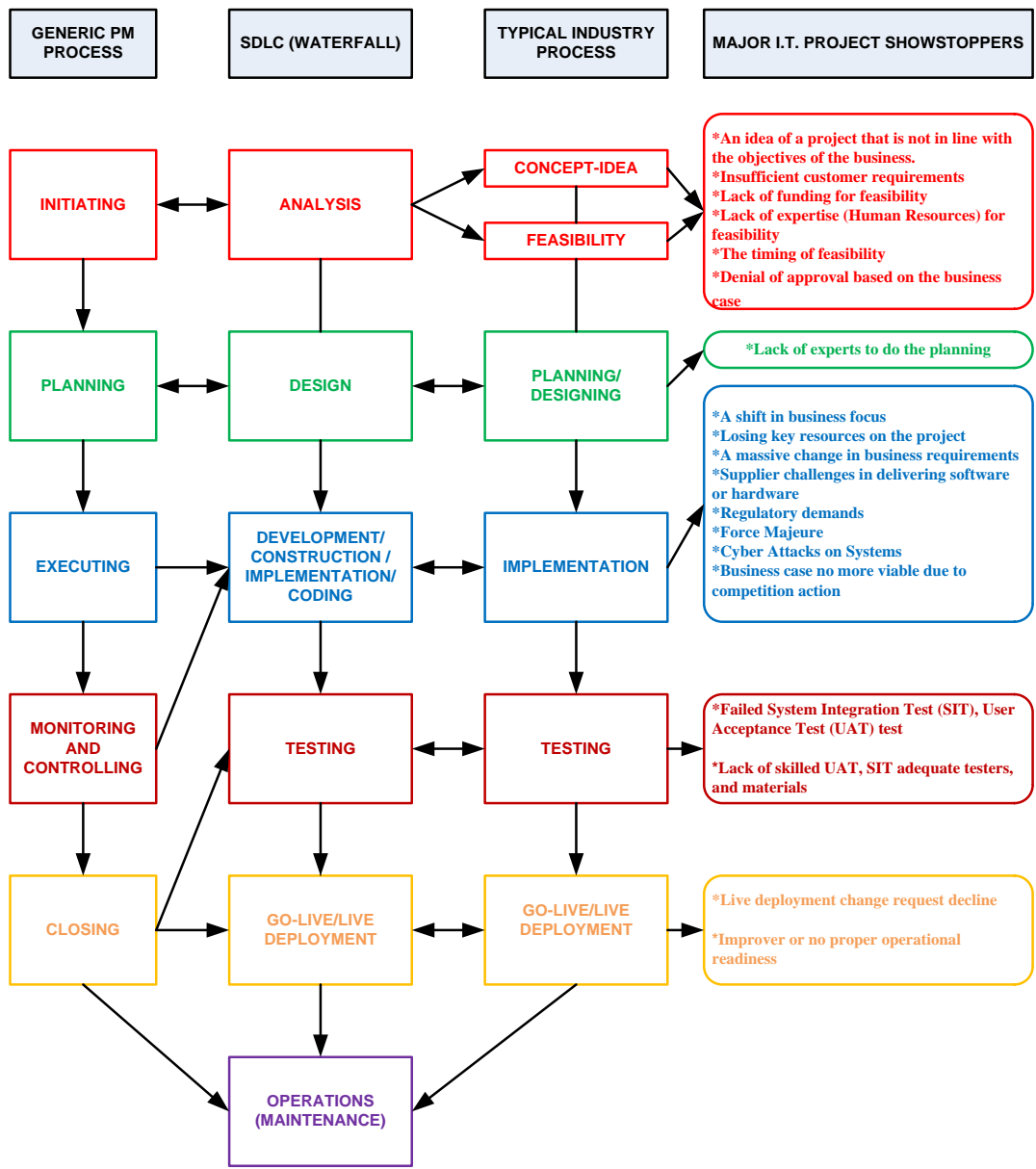

Figure 2: IT project showstopper framework (ITPSF)

## 5. CONCLUSION

From the discussions so far, one showstopper keeps appearing in all the stages, and that is "key human resource". The expertise and availability of project managers and team members are critical for the successful completion of IT projects. One other thing that featured prominently in the discussion was the fact that for an IT project to sail through from the concept phase to the live deployment phase, the project manager must be able to prove that the project is in line with the objectives of the business, the strategic direction of the business, is being mounted to gain competitive advantage, and has a solid business case. Thirdly, funding is key at all stages of the cycle for an IT project to be completed successfully. Lastly, but not least, approval for continuation at various stages is critical for a project to see the light of day. The output of this work can be used by project management professionals as a guide when delivering IT projects. It can serve as pointers for them (IT project managers) to quickly move in to resolve issues when they see some semblance with the showstoppers this work presents. An IT project manager for instance, should know that "insufficient customer requirements" can be a major showstopper to the progress or success of every IT project. With this knowledge, he/she can put in place measures right from the onset to forestall it from happening. The knack and in-depth knowledge of the project manager are key for the successful completion of an IT project. The output of this work presents a subset of the skill set an IT project manager needs to be able to function effectively and efficiently, especially in ensuring the successful completion of IT projects. Going forward, researchers in this area can employ quantitative means to explore the relationships between these showstoppers and IT project failure, to determine significant showstoppers, and so on.

**AUTHOR**

Godfred Yaw Koi-Akrofi is a senior lecturer and HOD for the Information Technology Studies Department of the University of Professional Studies, Accra in Ghana. He has Ph.D. in Management- Information Systems Management (ISM) from Universidad Central de Nicaragua (UCN), MBA in Management Information Systems (MIS) from University of Ghana Business School (UGBS), University of Ghana, Accra, Ghana, and Bachelor of Science (B.Sc.) in Electrical/Electronic Engineering, Kwame Nkrumah University of Science and Technology (KNUST), Kumasi, Ghana. Have worked with multinationals such as British Telecom, AT & T, MCI Communications Corp., Sprint Corporation, Verizon Communications, 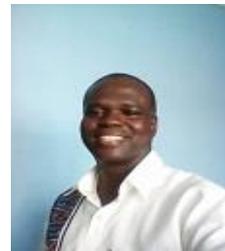
Deutsche Telekom, KPN, Orange, Telenor ASA, and so on as a Telecom engineer/Manager and Interconnect (Technical, Sales and Accounting) Manager with Ghana Telecom/Vodafone Ghana/Huawei Technologies for over 10 years. HaveTwelve years of experience as a lecturer in Tertiary education. Have 17 peer-reviewed journal articles, three books, and three conference papers. Areas of Research are in IT/IS Management, Telecommunications, MIS and the organization, Organizational Behaviour, and Management, Business Organizational Dynamics, Post-Merger/Acquisition Dynamics in the Telecoms and IT sectors, Quality Management, IS/IT/Business Strategy, IT/IS Project Management/ IT Design and Systems Thinking, IT investments and organizational performance, Complementary assets and value creation beyond IT investments.